\documentclass[journal]{IEEEtran}
\IEEEoverridecommandlockouts

\usepackage{cite}
\usepackage{comment}
\usepackage{amsmath,amssymb,amsfonts,bm}
\usepackage{nomencl}
\usepackage{etoolbox}
\usepackage{algorithm}
\usepackage{algorithmic}
\usepackage{graphicx}
\graphicspath{{Figures/}}
\usepackage{stfloats}
\usepackage{textcomp}
\usepackage[table]{xcolor}
\usepackage{hhline}
\usepackage{color,colortbl}
\usepackage{tikz}
\usepackage{multirow}
\usepackage{booktabs}
\usepackage{makecell}
\usepackage[hidelinks]{hyperref}
\usetikzlibrary{positioning,shapes.geometric,arrows.meta}

\usepackage{pgfplots}
\pgfplotsset{compat=1.17}
\definecolor{siZero}{RGB}{0,114,178}   
\definecolor{siOne}{RGB}{213,94,0}     
\definecolor{siThree}{RGB}{0,158,115}  

\definecolor{light-blue}{rgb}{0.6,0.6,1}
\definecolor{Gray}{gray}{0.9}

\def\BibTeX{{\rm B\kern-.05em{\sc i\kern-.025em b}\kern-.08em
T\kern-.1667em\lower.7ex\hbox{E}\kern-.125emX}}

\begin{document}

\title{Severity-Aware Minimum Security Sets for Accelerated Preventive SCOPF Using Machine Learning}

\author{\IEEEauthorblockN{ Mohammad Iman Alizadeh,
 Saeed Zeynali, Salah Ghamizi, Sylvain Kubler}
\thanks{This research was funded  by the Luxembourg National Research Fund (FNR) as part of the D2ET (Data-Driven Energy Transition) project, grant reference
 NCER25/SR/18324647}
}

\maketitle

\begin{abstract}
This paper proposes a severity-aware machine learning framework for reducing the computational burden of preventive AC security-constrained optimal power flow (PSCOPF). The central idea is to identify a compact minimum security set (MSS) containing only the non-dominated contingencies and associated post-contingency constraints that materially influence the preventive operating point while satisfying a prescribed severity tolerance. An offline procedure combines AC security assessment, severity-based ranking, non-dominated screening, and iterative reduced PSCOPF. After each reduced PSCOPF solution, exhaustive security assessment over all credible contingencies identifies additional contingency--constraint pairs whose residual severity exceeds the specified threshold. Severity is characterized using the normalized number, accumulated magnitude, maximum magnitude, and average magnitude of post-contingency violations. The resulting operating-point-dependent MSSs are used as labels for a supervised multi-label learning model that predicts the security elements required to construct the reduced PSCOPF directly from the pre-contingency state. The proposed framework therefore focuses the optimization on the contingencies and constraints that affect preventive control decisions, reducing problem size and computation time while explicitly controlling residual post-contingency severity. Results on the IEEE 60-bus and  118-bus systems, the latter having an approximately seven-times-larger PSCOPF formulation, demonstrate substantial computational reductions with limited impact on the preventive operating solution.
\end{abstract}

\begin{IEEEkeywords}
AC preventive security-constrained optimal power flow, contingency filtering, constraint screening, machine learning, minimum security set, security assessment, severity index.
\end{IEEEkeywords}

\section*{Nomenclature}

\begin{IEEEdescription}[\IEEEusemathlabelsep
\IEEEsetlabelwidth{$P_{G,g}^{c},Q_{G,g}^{c}$}]

\item[$\mathcal N$]
Set of buses, indexed by $i,j$.

\item[$\mathcal G,\mathcal G(i)$]
Set of generators and subset connected to bus $i$,
respectively; $g$ denotes a generator.

\item[$g_{\mathrm{sl}}$]
Slack-generator index.

\item[$\mathcal C$]
Set of credible contingencies, indexed by $c$.

\item[$\mathcal L^{c}$]
Set of  branches in state $c$, indexed by $l$.

\item[$f(\bm P_G^{0})$]
Pre-contingency generation cost.

\item[$P_{G,g}^{c},Q_{G,g}^{c}$]
Active and reactive outputs of generator $g$.

\item[$P_{D,i},Q_{D,i}$]
Active and reactive demands at bus $i$.

\item[$V_i^{c},\theta_i^{c}$]
Voltage magnitude and phase angle at bus $i$.

\item[$G_{ij}^{c},B_{ij}^{c}$]
Conductance and susceptance entries of the bus-admittance
matrix in state $c$.

\item[$P_l^{c},Q_l^{c}$]
Active and reactive power flows at the specified terminal
of branch $l$ in state $c$.

\item[$\overline F_l$]
Apparent-power limit of branch $l$.

\item[$x^{c}$]
Operating-state vector
$(\bm P_G^{c},\bm\theta^{c},\bm V^{c},\bm Q_G^{c})$.

\item[$(\cdot)^{\min},(\cdot)^{\max}$]
Lower and upper operating limits.

\end{IEEEdescription}

\noindent

\section{Introduction}

Preventive security-constrained optimal power flow (PSCOPF) is a fundamental tool for determining an economical operating point while maintaining the system within its operational limits following a set of credible contingencies \cite{stott2012optimal}. In its AC formulation, PSCOPF simultaneously represents the nonlinear pre-contingency operating state and the post-contingency states associated with the considered $N-1$ outages. Although this provides an accurate representation of network security, it also leads to a large-scale, nonlinear, and non-convex optimization problem whose computational burden increases rapidly with the size of the network and the number of credible contingencies \cite{florin_louis}.

Contingency filtering has long been recognized as an effective means of reducing the computational burden of PSCOPF \cite{capitanescu2007contingency,jiang2013novel,const_boufard}. A seminal contribution in this direction was presented in \cite{capitanescu2007contingency}, where an iterative PSCOPF framework was combined with contingency filtering based on the comparison of post-contingency constraint violations. The proposed filtering rules were shown to provide a more robust selection of critical contingencies than conventional severity-index-based ranking. This idea was subsequently extended to realistic large-scale nonlinear SCOPF in \cite{platbrood2014generic}, where contingency filtering was combined with network compression to reduce both the number and the complexity of the post-contingency models explicitly represented in the optimization problem.

A related stream of research has focused on removing redundant security constraints rather than only contingencies. For example, \cite{madani2016constraint} developed a constraint-screening methodology for identifying a small subset of security constraints sufficient to represent a much larger candidate set, while \cite{weinhold2020fast} combined low-impact and redundancy screening to construct a compact representation of security-constrained feasible injections. More recently, constraint filtering has also been extended to comprehensive AC-SCOPF formulations under uncertainty \cite{affine_iman}, demonstrating that a large fraction of contingency-dependent constraints can be safely eliminated before solving the nonlinear AC problem. These works confirm that substantial computational savings can be achieved not only by reducing the contingency set but also by exploiting the sparsity and redundancy of the associated post-contingency constraints. However, these approaches primarily focus on screening redundant or low-impact constraints from a predefined security representation, without explicitly considering the severity of the residual post-contingency violations. This highlights the need for screening strategies that jointly consider contingency--constraint relevance and violation severity, rather than relying solely on binary or redundancy-based filtering criteria.

Machine learning provides a complementary means of avoiding repeated screening calculations \cite{yan_hybrid,Hybrid_Data_Mode,popli2024robustness}.  In both \cite{elnaz_filtering} and \cite{popli2024robustness}, a contingency is classified as binding whenever at least one associated post-contingency inequality constraint has a non-zero dual multiplier. While this criterion is effective for avoiding false negatives, it is conservative because a contingency is retained even when only a single constraint is marginally active, and it does not distinguish the severity or multiplicity of the activated constraints. 

Despite the important developments, several limitations remain relevant to the construction of reduced AC-PSCOPF models. First, most contingency-filtering approaches operate primarily at the contingency level: once a contingency is retained, a comparatively large set of its associated post-contingency constraints may still be represented explicitly. Although dedicated constraint-screening methods exist \cite{madani2015promises}, simultaneous identification of both the relevant contingencies and their relevant post-contingency constraints has received considerably less attention, particularly in nonlinear AC-PSCOPF and ML-based screening frameworks.

Second, most existing filtering approaches are designed to reproduce the strict security requirements of the original PSCOPF. A contingency or constraint is therefore generally classified as either relevant or irrelevant according to whether it can become binding or violated \cite{elnaz_filtering}. Such a binary characterization does not provide a mechanism for controlling the degree of approximation of the reduced security model. In particular, a very small marginal violation and a severe post-contingency state may both cause a contingency or constraint to be retained, although their operational implications are substantially different, \cite{capitanescu2007contingency}.

The objective of PSCOPF reduction is therefore not only to identify which post-contingency states can become infeasible, but also to determine which security elements materially influence the preventive operating point \cite{preventinve}. From this perspective, the magnitude and structure of the residual violations are relevant in addition to their mere occurrence. A contingency producing a single marginal violation does not carry the same security information as one producing several large thermal or voltage violations. Similarly, contingencies producing largely overlapping violation patterns may be redundant even when both are individually classified as critical \cite{Redundancy}.

To address these needs, a severity-aware minimum security set (MSS) is introduced, comprising the minimum set of contingencies and associated inequality constraints required to satisfy a prescribed post-contingency severity level. The purpose of the severity-aware MSS is not merely to reduce the computational size of PSCOPF, but also to provide additional flexibility in preventive operation when limited post-contingency deviations can be operationally tolerated. Enforcing every nominal limit as a strict hard constraint may lead to unnecessarily conservative preventive actions, such as renewable curtailment or load shedding, even when the network can safely accommodate a small or temporary exceedance \cite{ROALD2023108725}. By distinguishing marginal from severe post-contingency violations, the proposed MSS allows the preventive operating point to exploit available network headroom while maintaining the residual security impact within a prescribed severity level.

A further motivation for the proposed MSS is that contingency criticality is strongly dependent on the pre-contingency operating point. The same network outage may be non-critical under one loading and generation pattern, but may become security-relevant under another due to changes in power flows, voltage profiles, reactive-power margins, and available control actions. Therefore, the security set should not be viewed as a fixed predefined list of contingencies, but rather as an operating-point-dependent representation of the contingencies and constraints that can materially influence the preventive operating decision \cite{velloso2021combining}. The proposed ML-based MSS captures this dependence by mapping the pre-contingency system state directly to the relevant contingency--constraint set.

In addition, non-dominance screening is used to avoid retaining multiple contingencies that convey essentially the same security information. A contingency is removed when its violated-constraint set and corresponding violation magnitudes are covered by a more severe contingency. The resulting MSS therefore operates at the contingency--constraint level and retains only the security information required for the prescribed severity tolerance.

The main contributions of this paper are as follows:
\begin{itemize}

\item A severity-aware security criterion is introduced for reduced AC-PSCOPF, combining the normalized cardinality, accumulated magnitude, maximum magnitude, and average magnitude of post-contingency violations. The prescribed severity tolerance provides the operator with an explicit means to balance computational complexity against residual post-contingency severity, avoiding unnecessarily conservative preventive actions when limited deviations can be operationally tolerated.

\item An operating-point-dependent contingency--constraint MSS construction procedure is developed by combining severity-based prioritization, non-dominated contingency screening, iterative reduced PSCOPF, and exhaustive AC security assessment. Rather than relying on a fixed predefined list of critical contingencies, the proposed MSS identifies, for each operating condition, the contingencies and individual post-contingency constraints that can materially influence the preventive operating point.

\item A supervised multi-label learning framework is developed to predict the severity-dependent MSS directly from the pre-contingency operating state. The ML model therefore learns how the relevant security set changes with system loading, generation, voltages, and power flows, and predicts the reduced security representation rather than the PSCOPF solution itself. This substantially reduces contingency-dependent state variables, nonlinear AC power-flow equations, and inequality constraints before solving the PSCOPF.

\end{itemize}

The remainder of this paper is organized as follows. Section~II formulates the full AC-PSCOPF problem and introduces the proposed severity-aware security criterion used to control residual post-contingency severity. Section~III presents the operating-point-dependent MSS construction procedure, including severity-based prioritization, non-dominance screening, iterative reduced PSCOPF, and exhaustive security assessment, followed by the supervised ML framework for MSS prediction. Section~IV evaluates the proposed method on two test systems with particular emphasis on reducing the problem size and computation time.
\section{Severity-Aware AC-PSCOPF}\label{Sev_JCC}

\subsection{Generic Preventive AC-PSCOPF Formulation}

Let $\mathcal{C}$ denote the set of credible contingencies and let
$c=0$ represent the pre-contingency operating state. The single-period
preventive AC-PSCOPF is formulated as
\begin{equation}
\min_{\bm P_G^{0}} \; f(\bm P_G^{0})
\label{eq:obj_pscopf}
\end{equation}
subject to the following constraints for all
$c\in\mathcal{C}\cup\{0\}$.

\begin{align}
&\sum_{g\in\mathcal{G}(i)} P_{G,g}^{c}
-P_{D,i}
=
V_i^{c}
\sum_{j\in\mathcal{N}}
V_j^{c}
\Big[
G_{ij}^{c}\cos(\theta_i^{c}-\theta_j^{c})
\nonumber\\
&
+B_{ij}^{c}\sin(\theta_i^{c}-\theta_j^{c})
\Big],
\hspace{1.5cm} \forall i\in\mathcal{N},
\ \forall c\in\mathcal{C}\cup\{0\},
\label{eq:acP_generic}
\\
&\sum_{g\in\mathcal{G}(i)} Q_{G,g}^{c}
-Q_{D,i}
=
V_i^{c}
\sum_{j\in\mathcal{N}}
V_j^{c}
\Big[
G_{ij}^{c}\sin(\theta_i^{c}-\theta_j^{c})
\nonumber\\
&
-B_{ij}^{c}\cos(\theta_i^{c}-\theta_j^{c})
\Big],
 \hspace{1.5cm} \forall i\in\mathcal{N},
\ \forall c\in\mathcal{C}\cup\{0\},
\label{eq:acQ_generic}
\\
&P_{G,g}^{\min}
\leq P_{G,g}^{c}
\leq P_{G,g}^{\max},
 \hspace{1.3cm} \forall g\in\mathcal{G},
\ \forall c\in\mathcal{C}\cup\{0\},
\label{eq:PGlim}
\\
&Q_{G,g}^{\min}
\leq Q_{G,g}^{c}
\leq Q_{G,g}^{\max},
\hspace{1.3cm} \forall g\in\mathcal{G},
\ \forall c\in\mathcal{C}\cup\{0\},
\label{eq:QGlim}
\\
&P_{G,g}^{c}
=
P_{G,g}^{0},
\hspace{2.5cm} \forall g\in\mathcal{G}\setminus\{g_{\mathrm{sl}}\},
\ \forall c\in\mathcal{C},
\label{eq:preventive_dispatch}
\\
&V_i^{\min}
\leq
V_i^{c}
\leq
V_i^{\max},
\hspace{1.6cm} \forall i\in\mathcal{N},
\ \forall c\in\mathcal{C}\cup\{0\},
\label{eq:Vlim_generic}
\\
&\left(P_l^{c}\right)^2+
\left(Q_l^{c}\right)^2
\leq
\left(\overline{F}_{l}\right)^2,
\hspace{0.9cm} \forall l\in\mathcal{L}^{c},
\ \forall c\in\mathcal{C}\cup\{0\}.
\label{eq:Flim_generic}
\end{align}

The contingency-dependent state vector is defined as
\begin{equation}
x^{c}
=
\left(
\bm{P_{G}}^{c},
\bm{\theta}^{c},
\bm{V}^{c},
\bm{Q}_{G}^{c}
\right),
\end{equation}
whereas the preventive active-power dispatch $\bm P_G^{0}$ is
determined before the occurrence of any contingency.

Equations~\eqref{eq:acP_generic}--\eqref{eq:acQ_generic} represent
the nonlinear AC active- and reactive-power balance equations in the
pre- and post-contingency states. Equations~\eqref{eq:PGlim} and
\eqref{eq:QGlim} enforce generator capability limits, while
\eqref{eq:Vlim_generic} and \eqref{eq:Flim_generic} impose bus-voltage
and branch thermal limits, respectively. The contingency-dependent
parameters $G_{ij}^{c}$, $B_{ij}^{c}$, and $\mathcal{L}^{c}$ represent
the network topology following contingency $c$.

The preventive nature of the formulation is therefore imposed on all
non-slack generators through \eqref{eq:preventive_dispatch}. Their
active-power schedules are fixed at the pre-contingency operating
point and remain unchanged following a contingency. The active power
of the slack generator is allowed to adapt to each post-contingency
state in order to compensate for the remaining active-power imbalance
and network losses. Reactive-power outputs may likewise vary within
their capability limits as the AC operating state changes.

The motivation for the proposed severity-aware MSS is not necessarily to obtain a cheaper preventive operating point, but rather to avoid the substantial computational effort required to eliminate security violations that have only a marginal impact on the resulting preventive solution. In the full PSCOPF, every represented post-contingency constraint must be satisfied irrespective of the magnitude or operational significance of its potential violation. Consequently, a small number of marginal violations may require the explicit representation of many contingency-dependent states and constraints, considerably increasing the optimization burden even when their removal has almost no effect on the optimal preventive dispatch.

The proposed MSS introduces a controlled severity tolerance into this process. Contingencies and constraints associated with significant post-contingency violations remain represented in the reduced PSCOPF, whereas marginal security elements need not enlarge the optimization problem when their residual severity remains below the prescribed level. The resulting objective is therefore to obtain a substantially smaller and faster preventive PSCOPF while preserving, as closely as possible, the economic operating point of the full formulation. The numerical results show that a very limited number of residual violations can be tolerated in exchange for substantial reductions in computation time, while the preventive objective value remains almost unchanged.

Because this reduction deliberately excludes part of the complete security representation from the optimization, exhaustive post-optimization security assessment becomes particularly important. Its role is to quantify the residual violations over the complete contingency set and to verify that the computational savings obtained through the MSS do not result in post-contingency severity beyond the prescribed tolerance.
\subsection{Severity-Aware Risk Extension}

To characterize the operational impact of post-contingency
violations, four normalized severity indices are evaluated
for each contingency $c\in\mathcal C$ under a given
operating point sample $s\in\mathcal S$:
\begin{align}
\mathrm{SI}_{1}^{(s)}(c) &= \text{normalized total number of constraint violations},\\
\mathrm{SI}_{2}^{(s)}(c) &= \text{normalized sum of constraint violations},\\
\mathrm{SI}_{3}^{(s)}(c) &= \text{normalized maximum constraint violation},\\
\mathrm{SI}_{4}^{(s)}(c) &= \text{normalized average constraint violation}.
\end{align}
These complementary indices jointly characterize both
the number and magnitude of post-contingency violations.
Their weighted aggregation defines the severity index
for contingency $c$ under operating point sample $s$:
\begin{equation}
\mathcal{S}^{(s)}(c)=
\sum_{\ell=1}^{4} w_\ell \mathrm{SI}_{\ell}^{(s)}(c),
\label{eq:severity}
\end{equation}
where $w_\ell$ denotes the weighting coefficient associated
with the $\ell$th severity component.

The severity index provides a quantitative measure of
the operational impact of each contingency, distinguishing
between marginal and severe operating-limit violations.
A prescribed severity threshold $\beta$ is used to identify
contingencies requiring further consideration during
the filtering procedure:
\begin{equation}
\mathcal{S}^{(s)}(c)>\beta,
\qquad c\in\mathcal C,\;s\in\mathcal S.
\label{eq:severityrisk}
\end{equation}
Contingencies exceeding this threshold are considered relevant for further security assessment and constraint selection, while those below the threshold may be excluded from the reduced PSCOPF formulation, provided they are not subsequently identified as problematic during further security evaluations.
\section{Minimum Security Set Construction and ML Approximation}

Directly incorporating the proposed severity-aware risk criterion into AC-PSCOPF is computationally intractable for two reasons. First, the large number of contingency-dependent inequality constraints leads to a prohibitively large optimization problem. Second, the proposed severity indices are available only after solving the AC security assessment (SA) problem, since they are functions of the resulting post-contingency constraint violations rather than explicit optimization variables. Consequently, they cannot be directly embedded into the optimization model.

To overcome these limitations, the proposed framework does not approximate the PSCOPF solution itself. Instead, it learns the MSS, i.e., the minimum contingency--constraint representation required to satisfy the severity-aware risk criterion. The resulting reduced formulation preserves the AC-feasible operating point of the original problem while substantially decreasing the number of explicitly modelled contingencies and security constraints, thereby improving the tractability of severity-aware AC-PSCOPF.

\subsection{Overview}

For each operating point sample, the offline procedure first constructs an MSS using AC OPF, SA, severity-indexed screening, non-dominated contingency selection, and iterative reduced PSCOPF. The resulting MSS is then used as the label of a supervised multi-label learning problem. During online operation, the trained ML model receives the pre-contingency operating state as input and predicts the corresponding MSS.

The ML model learns the mapping
\begin{equation}
\mathbf{x}^{(s)}\longmapsto\mathcal M^{(s)},
\qquad \forall s\in\mathcal S,
\label{eq:sample_mapping}
\end{equation}
where $\mathbf{x}^{(s)}$ denotes the feature vector of  operating samples $s$, and $\mathcal M^{(s)}$ denotes the corresponding MSS. By regrouping the predicted MSSs according to their contingency index, the sample-wise representations are transformed into contingency-wise representations,
\begin{equation}
\left\{\mathcal M^{(s)}\right\}_{s\in\mathcal S}
\Longrightarrow
\left\{\left(c,\mathcal S_c,\mathcal K_c^{(s)}\right):c\in\mathcal C\right\},
\label{eq:regroup}
\end{equation}
where $\mathcal S_c\subseteq\mathcal S$ denotes the subset of  operating samples in which contingency $c$ belongs to the MSS, and $\mathcal K_c^{(s)}$ denotes the active subset of post-contingency inequality constraints associated with contingency $c$ under operating point $s$. 


\subsection{Initial Screening}

For each  operating sample $s\in\mathcal S$, the corresponding net-load profile, i.e., the load not supplied by RES, is first used to solve a base-case AC OPF. This AC OPF can be interpreted as a special case of PSCOPF without contingency constraints and is formulated as $c=0$
\begin{align}
&\min \hspace{0.5em}\eqref{eq:obj_pscopf}
\hspace{1em}\text{s.t.}\hspace{0.5em}\eqref{eq:acP_generic}-\eqref{eq:Flim_generic}.
\label{eq:opf}
\end{align}
The resulting system vector
\begin{equation}
\mathbf{x}^{(s)}=
[P_D^{(s)}, Q_D^{(s)}, P_G^{(s)}, Q_G^{(s)},
P_F^{(s)}, Q_F^{(s)}, V^{(s)}, \theta^{(s)}]
\label{eq:features}
\end{equation}
constitutes the feature vector representing the pre-contingency operating state, where $P_F^{(s)}$ and $Q_F^{(s)}$ denote active and reactive branch flows.


Algorithm~\ref{alg:initial_screening} constructs an initial set of 
security-relevant contingencies for each operating sample  $  x^{(s)} \in \mathbf{x}^{(s)}$. This operating point is 
then subjected to an exhaustive SA over the credible 
contingency set $\mathcal C$. For each contingency $c$, the SA returns the vector $v^{(s)}(c)$, whose element 
$v_k^{(s)}(c)$ quantifies the violation of operational constraint $k$. 
Accordingly,
\begin{equation}
\mathcal K^{(s)}(c)
=
\left\{
k\in\mathcal N : v_k^{(s)}(c)>0
\right\}
\end{equation}
collects all constraints violated under contingency $c$.

The violation pattern for each contingency is subsequently characterized 
by the aggregate severity score $S^{(s)}(c)$
 introduced in Section~\ref{Sev_JCC}.

The contingencies are then ranked in decreasing order of $S^{(s)}(c)$, 
such that contingencies producing more severe post-contingency states 
receive higher priority. The resulting ordered set is denoted by 
$\mathcal C_{\mathrm{rank}}^{(s)}$ and is subsequently subjected to 
non-dominance screening.

The non-dominance screening removes contingencies that provide no 
additional security information relative to another contingency in the 
ranked set. Specifically, contingency $c$ is considered dominated by 
$c'$ if
\begin{equation}
\mathcal K^{(s)}(c)
\subseteq
\mathcal K^{(s)}(c')
\quad \text{and} \quad
S^{(s)}(c)
\leq
S^{(s)}(c').
\label{eq:contingency_dominance}
\end{equation}
The first condition indicates that every constraint violated by $c$ is 
also violated by $c'$, while the second ensures that $c'$ has at least 
the same aggregate severity. Hence, retaining $c$ would introduce 
neither a new violated constraint nor a more severe security condition. 
Only contingencies for which no such dominating contingency exists are 
included in $\mathcal C_{\mathrm{cand}}^{(s)}$. The resulting 
non-dominated candidate set provides a compact, severity-representative 
initialization for the subsequent iterative MSS construction.

\begin{algorithm}[tb]
\caption{Severity-based initial screening with non-dominated contingency selection}
\label{alg:initial_screening}
\begin{algorithmic}[1]
\REQUIRE  Operating point $s\in\mathcal S$ with net-load profile $(P_D^{(s)},Q_D^{(s)})$; credible contingency set $\mathcal C$; severity weights $w_\ell$, $\ell=1,\ldots,4$
\ENSURE Initial non-dominated candidate contingency set $\mathcal C_{\mathrm{cand}}^{(s)}$
\STATE Solve the base-case AC OPF:
\\\vspace{3pt}
$x^{(s)}\gets \mathrm{OPF}\!\left(P_D^{(s)},Q_D^{(s)}\right)$
\vspace{3pt}

\STATE Initialize $\mathcal C_{\mathrm{rank}}^{(s)}\gets\emptyset$ and $\mathcal C_{\mathrm{cand}}^{(s)}\gets\emptyset$
\FOR{each contingency $c\in\mathcal C$}
    \STATE Run AC security assessment:
    \\\vspace{3pt}
    $v^{(s)}(c)\gets \mathrm{SA}\!\left(x^{(s)},c\right)$
    \vspace{3pt}
    \STATE Extract the violated constraint set:
    \\\vspace{3pt}
    $\mathcal K^{(s)}(c)\gets\{k\in\mathcal N:\;v_k^{(s)}(c)>0\}$
    \vspace{3pt}
    \STATE Compute $\mathrm{SI}_{\ell}^{(s)}(c)$, $\ell=1,\ldots,4$
    \STATE Compute the aggregate severity score:
\\\vspace{3pt}
    $S^{(s)}(c)\gets \sum_{\ell=1}^{4} w_\ell \mathrm{SI}_{\ell}^{(s)}(c)$
    \vspace{3pt}
\ENDFOR
\STATE Rank contingencies in decreasing order of $S^{(s)}(c)$
\STATE Select the top-ranked contingencies:
\\\vspace{3pt}
$\mathcal C_{\mathrm{rank}}^{(s)}
\gets \operatorname{SelectTop}\!\left(\mathcal C,S^{(s)}\right)$
\vspace{3pt}
\FOR{each contingency $c\in\mathcal C_{\mathrm{rank}}^{(s)}$}
    \STATE $\mathrm{dominated}\gets\mathrm{false}$
    \FOR{each contingency $c'\in\mathcal C_{\mathrm{rank}}^{(s)}$, $c'\neq c$}
        \IF{$\mathcal K^{(s)}(c)\subseteq \mathcal K^{(s)}(c')$ \textbf{and} $S^{(s)}(c)\le S^{(s)}(c')$}
            \STATE $\mathrm{dominated}\gets\mathrm{true}$
        \ENDIF
    \ENDFOR
    \IF{$\mathrm{dominated}=\mathrm{false}$}
        \STATE $\mathcal C_{\mathrm{cand}}^{(s)}\gets \mathcal C_{\mathrm{cand}}^{(s)}\cup\{c\}$
    \ENDIF
\ENDFOR
\STATE \textbf{return} $\mathcal C_{\mathrm{cand}}^{(s)}$
\end{algorithmic}
\end{algorithm}

\subsection{Iterative Construction of the MSS}

Algorithm~\ref{alg:iterative_mss} refines the candidate set obtained from 
Algorithm~\ref{alg:initial_screening} to construct the MSS associated with 
 operating sample $s$. The procedure is initialized with the 
contingency--constraint pairs identified during the initial security 
assessment. Specifically,
\begin{equation}
\mathcal M^{(s)}
=
\left\{
(c,k):
c\in\mathcal C_{\mathrm{cand}}^{(s)},
\;
k\in\mathcal K^{(s)}(c)
\right\},
\label{eq:mss_initialization}
\end{equation}
where each pair $(c,k)$ indicates that post-contingency constraint $k$ 
is explicitly enforced for contingency $c$. This initialization therefore 
contains the violated constraints associated with the non-dominated 
contingencies identified by the initial screening procedure.

At iteration $r$, a reduced PSCOPF is solved by explicitly enforcing only 
the contingency--constraint pairs contained in $\mathcal M^{(s)}$. The 
resulting preventive operating point is subsequently subjected to a complete 
SA over the original contingency set $\mathcal C$. 
Importantly, this assessment is not restricted to the contingencies or 
constraints currently included in $\mathcal M^{(s)}$ and therefore verifies 
the reduced PSCOPF solution against the complete security model.
Based on the updated aggregate severity score $S^{(s)}(c)$, the residual security impact is assessed after enforcing the constraints currently included in $\mathcal M^{(s)}$ through the reduced PSCOPF.

The complete SA may reveal violated constraints that are 
not yet represented in the current MSS. These unresolved 
contingency--constraint pairs are collected as
\begin{equation}
\Delta\mathcal M^{(s)}
=
\left\{
(c,k):
h_k^{c,s}>0,\;
(c,k)\notin\mathcal M^{(s)}
\right\},
\label{eq:unresolved_pairs}
\end{equation}
where $h_k^{c,s}>0$ indicates that constraint $k$ remains violated under 
contingency $c$ after solving the current reduced PSCOPF.

If  $  \mathcal S^{(s)}(c)\leq\beta$
the prescribed severity budget is satisfied over the complete contingency 
set and no further security constraints are required for the considered 
risk level. The iterative procedure therefore terminates and the current 
$\mathcal M^{(s)}$ is retained as the MSS.

Otherwise, the unresolved pairs in $\Delta\mathcal M^{(s)}$ are prioritized 
according to their violation magnitude and their contribution to the 
remaining contingency severity. The most critical pairs are added to 
$\mathcal M^{(s)}$, and the reduced PSCOPF is solved again. This 
constraint-generation process continues until the severity criterion is satisfied, no unresolved violated pairs remain, 
or the maximum number of iterations $N_{\max}$ is reached. Consequently, 
the procedure progressively introduces only those post-contingency 
constraints required to reduce the residual severity to the prescribed 
level, rather than imposing the complete set of security constraints from 
the outset.

\begin{algorithm}[t!]
\caption{Iterative construction of the minimum security set}
\label{alg:iterative_mss}
\begin{algorithmic}[1]
\REQUIRE Operating point $s\in\mathcal S$; candidate contingency set $\mathcal C_{\mathrm{cand}}^{(s)}$; violated constraint sets $\mathcal K^{(s)}(c)$; severity budget $\beta$; maximum iteration number $N_{\max}$
\ENSURE Minimum security set $\mathcal M^{(s)}$
\STATE Initialize $r\gets0$
\STATE Initialize
\\\vspace{3pt}
$\mathcal M^{(s)}\gets\{(c,k):c\in\mathcal C_{\mathrm{cand}}^{(s)},\;k\in\mathcal K^{(s)}(c)\}$
\vspace{3pt}
\REPEAT
    \STATE $r\gets r+1$
    \STATE Solve the reduced PSCOPF using only the pairs in $\mathcal M^{(s)}$
    \STATE Run SA for all $c\in\mathcal C$ using the reduced PSCOPF solution
    \STATE Compute updated severity scores:
    \\\vspace{3pt}
    $\mathcal S^{(s)}(c)=\sum_{\ell=1}^{4}w_\ell\mathrm{SI}_{\ell}^{(s)}(c),
    \qquad \forall c\in\mathcal C$
    \vspace{3pt}
    \STATE Identify unresolved violated pairs:
    \\\vspace{3pt}
    $\Delta\mathcal M^{(s)}\gets
    \{(c,k):h_k^{c,s}>0,\;(c,k)\notin\mathcal M^{(s)}\}$
    \vspace{3pt}
    \IF{$\mathcal S^{(s)}(c)\le\beta,\;\forall c\in\mathcal C$}
        \STATE \textbf{break}
    \ENDIF
    \STATE Rank $\Delta\mathcal M^{(s)}$ according to violation magnitude and severity contribution
    \STATE Add the most critical violated pairs:
    \\\vspace{3pt}
   $ \mathcal M^{(s)}\gets
    \mathcal M^{(s)}\cup \operatorname{SelectTop}\!\left(\Delta\mathcal M^{(s)}\right)$
    \vspace{3pt}
\UNTIL{$\Delta\mathcal M^{(s)}=\emptyset$ \textbf{or} $r=N_{\max}$}
\STATE \textbf{return} $\mathcal M^{(s)}$
\end{algorithmic}
\end{algorithm}

The final MSS is used as the target label for supervised learning. Since the reduced PSCOPF grows only when the full SA identifies unresolved violations, unnecessary contingencies and constraints are avoided while preserving the AC feasibility and severity-aware risk characteristics of the original problem.

\subsection{{Two-Stage Graph Attention Screening Model}}

{The map from a pre-contingency operating point to its MSS is, by nature, a problem defined \emph{on} the power network: whether a contingency becomes binding, and which limits it activates, depends on how injections and flows redistribute across neighbouring buses and branches once an element is lost. We therefore realise the learned mapping $f_{\rm ML}$ with a two-stage graph attention network (GAT), whose computational graph mirrors the electrical grid, i.e. buses are nodes and branches are edges. Because a GAT exchanges messages only along physically existing branches and lets every bus weight its neighbours through learned attention coefficients, it reproduces the local, topology-driven way in which post-contingency stress propagates far more naturally than a flat multilayer perceptron that discards connectivity. This inductive bias makes the model particularly well suited to security screening, where the binding pattern is strongly shaped by network structure.}

{The input is the pre-contingency state returned by the base-case AC OPF. Rather than flattening the system vector $\mathbf{x}^{(s)}$ of \eqref{eq:features} into a single array, we distribute it over the graph: each bus $i$ in the bus set $\mathcal V$ carries a node-feature vector collecting its own demand, generation, voltage magnitude and angle together with the active and reactive flows on its incident branches,}
\begin{equation}
\mathbf{x}_{i}^{(s)}=\big[\,P_{D,i},\,Q_{D,i},\,P_{G,i},\,Q_{G,i},\,V_{i},\,\theta_{i},\,\bar P_{F,i},\,\bar Q_{F,i}\,\big],\;\; i\in\mathcal V,
\label{eq:node_features}
\end{equation}
{where $\bar P_{F,i}$ and $\bar Q_{F,i}$ aggregate the incident branch flows and the edge set reproduces the branch topology of the test system. Node features are standardised using training-set statistics only.}

{A shared trunk of stacked multi-head graph-attention layers refines these node embeddings, and a permutation-invariant mean readout compresses them into a single sample embedding $\mathbf{h}^{(s)}$ that summarizes the complete operating point. Two lightweight heads then decode $\mathbf{h}^{(s)}$ into the MSS in the same order in which the offline procedure builds it. The \emph{Stage-A} head is a contingency classifier that returns a binding probability for each of the $N_C$ candidate contingencies,}
\begin{equation}
p^{A}_{c}=\operatorname{sig}\!\big(\mathbf{w}_{c}^{\top}\mathbf{h}^{(s)}+b_{c}\big),\qquad c\in\mathcal C,
\label{eq:stageA}
\end{equation}
{so that the predicted binding-contingency set is $\widehat{\mathcal C}^{(s)}=\{c:\,p^{A}_{c}>\tau_{A}\}$. Conditioned on the same embedding and on a learned embedding $\mathbf{e}_{c}$ of a retained contingency, the \emph{Stage-B} head returns, for every monitored limit, the probability that it is a binding post-contingency constraint,}
\begin{equation}
p^{B}_{c,k}=\operatorname{sig}\!\big(\phi_{k}(\mathbf{h}^{(s)},\mathbf{e}_{c})\big),\;\; k\in\{1,\dots,2N_{A}+2N_{B}\},
\label{eq:stageB}
\end{equation}
{covering the branch flow (i.e. both flows to and from the bus), and bus voltage lower/upper bounds, with predicted set $\{k:\,p^{B}_{c,k}>\tau_{B}\}$.}

{Supervision is exactly the multi-hot encoding of the MSS: the Stage-A target $y^{A}_{c}\in\{0,1\}$ marks the contingencies contained in $\mathcal M^{(s)}$, while the Stage-B target $y^{B}_{c,k}\in\{0,1\}$ marks the individual limits activated under each retained contingency. Both heads are trained jointly by minimizing a binary cross-entropy loss summed over the two stages, $\mathcal L=\mathcal L_{A}+\lambda_{B}\mathcal L_{B}$, so that a single forward pass yields the complete contingency--constraint prediction}
\begin{equation}
f_{\rm ML}:\;\mathbf{x}^{(s)}\longmapsto\mathcal M^{(s)}=\big\{(c,k):\,p^{A}_{c}>\tau_{A}\;\wedge\;p^{B}_{c,k}>\tau_{B}\big\}.
\label{eq:ml_mapping}
\end{equation}
{The two decision thresholds $(\tau_{A},\tau_{B})$ act as tunable operating points that trade coverage of the true MSS against the size of the predicted set, as analyzed in Section~IV.}

\section{Numerical Results}

The proposed methodology is evaluated on two test systems:
the IEEE 60-bus system, comprising 60 buses, 88
transmission lines and 41 credible contingencies \cite{nordic32}, and the IEEE 118-bus
system, comprising 118 buses,  185 transmission lines
and 140 plausible contingencies\cite{uw_pstca_1999}. The comprehensive performance
assessment is conducted on the IEEE 60-bus system,
while the IEEE 118-bus system is primarily used to
demonstrate the computational efficiency and scalability
of the proposed methodology on a larger PSCOPF problem. The larger system leads to a substantially larger PSCOPF formulation, approximately seven times larger in terms of the average number of variables and constraints in the full PSCOPF, as well as a considerably broader contingency set.

Credible transmission-line contingencies are selected by excluding those that result in network islanding, bus disconnection, or conflicting contingency pairs. The operating point samples are generated by allowing load deviations of up to 20\% at each bus in a uniform probability distribution function. The optimization problems are implemented in Julia, and the resulting nonlinear programming (NLP) problems are solved using IPOPT, \cite{wachter2002ipopt}.

To investigate the impact of the severity threshold on the proposed reduction framework, three different values of the severity index, namely $\beta = 0.0$, $\beta=0.1$, and $\beta=0.3$, are considered. Smaller values of $\beta$ retain contingencies and constraints associated with even minor violations, whereas larger values focus only on the most critical violations.

Detailed numerical results, including contingency reduction, constraint reduction, prediction accuracy, severity preservation, and computational time, are reported in the following subsections.

\subsection{Comparison with the Full PSCOPF Benchmark}

\begin{table}[!t]
\centering
\caption{Average performance of the reduced PSCOPF for IEEE 60-bus test system under different severity thresholds over 100 operating samples.}
\label{tab:severity_comparison}

\renewcommand{\arraystretch}{1.12}
\setlength{\tabcolsep}{5pt}

\begin{tabular}{l|ccc}
\hline
\textbf{Metric}
&
\textbf{Severity 0.0}
&
\textbf{Severity 0.1}
&
\textbf{Severity 0.3}
\\
\hline

\textbf{Full PSCOPF} \\
\cline{1-1}

Elapsed time (s)
& 6.3 & 6.3 & 6.3 \\

No. variables
& 50,948 & 50,948 & 50,948 \\

No. constraints
& 65,067 & 65,067 & 65,067 \\

\hline

\textbf{Reduced PSCOPF} \\
\cline{1-1}

Objective deviation (\%)
& 0
& 0
& 0 \\

Elapsed time (s)
& 1.5 & 1.2 & 1.0 \\
No. contingencies
& 18 & 18 & 14 \\

No. variables
& 23,039 & 20,185 & 16,975 \\

No. constraints
& 25,042 & 21,981 & 18,531 \\

\hline

\textbf{SA of reduced PSCOPF} \\
\cline{1-1}

Maximum violation
& 0
& 3.5$\times10^{-4}$
& 1.5$\times10^{-3}$ \\

Average violation
& 0
& 3.5$\times10^{-4}$
& 1.5$\times10^{-3}$ \\

No. violations
& 0
& 0.25
& 0.6 \\

Sum of violations
& 0
& 3.5$\times10^{-4}$
& 1.5$\times10^{-3}$ \\

\hline

\textbf{Overall Performance} \\
\cline{1-1}

Time reduction (\%)
& 75.0 & 80.0 & 84.1 \\

Contin. reduction (\%)
& 56 & 56 & 65.8 \\

Variable reduction (\%)
& 54.7 & 60.3 & 66.6 \\

Constraint reduction (\%)
& 61.5 & 66.2 & 71.5 \\

\hline
\end{tabular}
\end{table}

Table~\ref{tab:severity_comparison} compares the full and reduced PSCOPF for the three severity thresholds on the IEEE 60-bus system. All cases are evaluated using the same 100 operating samples, enabling a direct comparison across severity levels.

The reduced PSCOPF provides substantial computational savings. The average solution time decreases from approximately $6$~s for the full PSCOPF to $1.5$, $1.2$, and $1.0$~s for severity thresholds $0.0$, $0.1$, and $0.3$, corresponding to reductions of $75\%$, $81\%$, and $82\%$, respectively. Meanwhile, the objective deviation remains unchanged.

The computational gains are consistent with the reduction in problem size, mainly driven by the reduction in the number of retained contingencies across all severity levels, ranging from $56\%$ at severity $0.0$ to $65.8\%$ at severity $0.3$. Relative to the full PSCOPF with $50{,}948$ variables and $65{,}067$ constraints, the variable reduction increases from $54.7\%$ to $66.6\%$, while the constraint reduction increases from $61.5\%$ to $71.5\%$. Hence, increasing the severity threshold progressively removes marginal contingency--constraint pairs and produces a more compact optimization problem.

The post-optimization security assessment shows the corresponding security trade-off. No residual violation is observed at severity $0.0$, whereas severity $0.1$ results in only $0.25$ violations per sample on average, with an average magnitude of $3.5\times10^{-4}$. At severity $0.3$, these values increase to $0.62$ violations and $1.5\times10^{-3}$, respectively. Thus, higher severity thresholds provide additional computational savings at the cost of a small increase in residual post-contingency violations.

\subsection{Performance on a Larger Test System}

To further assess the scalability of the proposed framework, additional simulations are performed on the IEEE
118-bus test system.  It therefore provides a more demanding benchmark for evaluating the computational benefits of the proposed MSS-based reduction. The credible contingency
set contains $140$ transmission-line contingencies, obtained after
excluding islanding, infeasible, and conflicting contingencies from the
initial outage set. For the 100-sample experiments, the reduced
security representation retains on average only $85$ contingencies,
showing that a significant fraction of the original contingency set
does not need to be represented explicitly in each reduced PSCOPF.

Table~\ref{tab:severity_comparison_118} shows that the proposed reduction provides substantial computational savings for the IEEE 118-bus system. The full PSCOPF contains $356{,}688$ variables and $456{,}805$ constraints and requires approximately $152$~s to solve. The reduced PSCOPF decreases the solution time to $30$, $21$, and $16$~s for severity levels $0.0$, $0.1$, and $0.3$, corresponding to time reductions of $80.3\%$, $86.2\%$, and $89.5\%$, respectively.

A main driver of this improvement is the progressive reduction in the number of contingencies represented in the PSCOPF. From the full set of $140$ credible contingencies, the reduced formulation retains $128$, $69$, and $59$ contingencies as the severity threshold increases, corresponding to reductions of $8.5\%$, $50.7\%$, and $57.8\%$. This directly reduces the number of contingency-dependent variables and constraints. Consequently, the variable reduction increases from $9.3\%$ to $39.2\%$, while the constraint reduction increases from $23.1\%$ to $48.4\%$.

These computational gains are obtained with essentially no change in the preventive objective. The post-optimization SA, however, shows the expected security trade-off: no residual violation is observed at severity $0.0$, whereas severity levels $0.1$ and $0.3$ result in an average of two residual violations, with maximum violation values of $0.92$ and $0.95$, respectively. 
\begin{table}[!t]
\centering
\caption{Average performance of the reduced PSCOPF for IEEE 118-bus test system under different severity thresholds over 100 operating samples.}
\label{tab:severity_comparison_118}

\renewcommand{\arraystretch}{1.12}
\setlength{\tabcolsep}{5pt}

\begin{tabular}{l|ccc}
\hline
\textbf{Metric}
&
\textbf{Severity 0.0}
&
\textbf{Severity 0.1}
&
\textbf{Severity 0.3}
\\
\hline

\textbf{Full PSCOPF} \\
\cline{1-1}

Elapsed time (s)
& 152 & 152 & 152 \\

No. variables
& 356,688 & 356,688 & 356,688 \\

No. constraints
& 456,805 & 456,805 & 456,805 \\

\hline

\textbf{Reduced PSCOPF} \\
\cline{1-1}

Objective deviation (\%)
& 0.0 & 0.0 & 0.0 \\

Elapsed time (s)
& 30 & 21 & 16 \\

No. contingencies
& 128 & 69 & 59 \\
No. variables
& 323,668 & 222,068 & 216,988 \\

No. constraints
& 351,486 & 241,280 & 235,779 \\

\hline

\textbf{SA of reduced PSCOPF} \\
\cline{1-1}

Maximum violation
& 0.0 & 0.92 & 0.95 \\

Average violation
& 0.0 & 0.92 & 0.95 \\

No. violations
& 0.0 & 2.0 & 2.0 \\

Sum of violations
& 0.0 & 0.92 & 0.95 \\

\hline

\textbf{Overall Performance} \\
\cline{1-1}

Time reduction (\%)
& 80.3 & 86.2 & 89.5 \\

Contin. reduction (\%)
& 8.5 & 50.7 & 57.8 \\

Variable reduction (\%)
& 9.3 & 37.7 & 39.2 \\

Constraint reduction (\%)
& 23.1 & 47.2 & 48.4 \\

\hline
\end{tabular}
\end{table}

Overall, the results indicate that the severity threshold primarily
controls a trade-off between computational burden and residual
post-contingency severity rather than between economic performance and
security. The objective value remains essentially unchanged, whereas
the number of security elements embedded in the optimization and the
corresponding solution time decrease considerably. The exhaustive
post-optimization security assessment is therefore an essential part of
the proposed framework: the reduced PSCOPF determines the preventive
operating point using a compact security representation, while the full
security assessment evaluates all omitted contingency--constraint pairs
and quantifies any residual violations. This separation enables a
substantial reduction in PSCOPF computational burden while preserving
the preventive economic solution and maintaining explicit visibility
over the complete contingency set.

It should also be noted that the final results averaged over multiple samples do not necessarily reproduce the targeted severity threshold exactly. First, the severity threshold is applied to individual contingency violations during the filtering process, whereas the reported severity metrics represent aggregated performance over the complete set of contingencies and operating samples. Second, due to the nonlinear nature of the PSCOPF, the final preventive operating point does not scale proportionally with the imposed severity threshold. A small change in the retained contingency--constraint set may shift the preventive operating point toward a more conservative solution, resulting in residual violations smaller than the prescribed threshold. This behavior is observed in both test systems.

Nevertheless, the proposed framework retains a direct corrective mechanism through the post-optimization security assessment. If the SA identifies a residual violation whose severity exceeds the prescribed threshold, the corresponding contingency and violated constraints can simply be added to the reduced PSCOPF and the problem re-solved. This corrective step requires no modification of the PSCOPF formulation and can be repeated until all residual violations satisfy the prescribed severity threshold.
\subsection{Sparsity and recurrence of retained contingency sets}


Let $\mathcal{C}$ denote the set of 41 credible contingencies and
$B_i\subseteq\mathcal{C}$ the retained set for operating point $i$,
with $N=5{,}000$. The selection sparsity is defined as
$1-\sum_{i=1}^{N}|B_i|/(N|\mathcal{C}|)$.
Table~\ref{tab:selection-sparsity} shows that only 16.08 contingencies
are retained on average, corresponding to a sparsity of 60.77\%.
Moreover, the 5,000 distinct load profiles produce only 27 distinct
retained sets, and the five most frequent sets account for 76.12\%
of the samples. This combination of sparsity and recurrence motivates
learning the relationship between operating conditions and contingency
selection.

\begin{table}[htbp]
\centering
\small
\caption{Sparsity and recurrence of contingency selection across 5,000 IEEE 60-bus operating points.}
\label{tab:selection-sparsity}
\renewcommand{\arraystretch}{1.12}
\begin{tabular}{@{}lr@{}}
\toprule
Statistic & Value \\
\midrule
Credible contingencies, $|\mathcal{C}|$ & 41 \\
Mean retained contingencies & 16.08 \\
Mean omitted contingencies & 24.92 \\
Selection sparsity & 60.77\% \\
Range of retained-set sizes & 14--19 \\
\midrule
Contingencies retained in every sample & 12 \\
Contingencies retained in some but not all samples & 8 \\
Contingencies never retained in the dataset & 21 \\
\midrule
Distinct retained sets & 27 \\
Samples covered by the five most frequent sets & 76.12\% \\
\bottomrule
\end{tabular}
\par\smallskip
\end{table}

The 5,000-point label export further exhibits a persistent core and a rare tail: twelve contingency IDs are retained in every operating point, whereas ID 27 appears only three times. A static set of the 15 most frequent labels covers 91.30\% of label events but contains the complete exported set in only 35.58\% of operating points. Thus, aggregate label coverage should be complemented by operating-point-level complete-set coverage and post-optimization security assessment. Coverage is $\sum_i|B_i\cap C_k|/\sum_i|B_i|$; complete OPs (i.e. operating points) satisfy $B_i\subseteq C_k$. There are 80,413 label events. Ties are resolved by ascending ID. The 20 observed IDs are not the complete credible-contingency universe.

\begin{table}[htbp]
\centering
\footnotesize
\setlength{\tabcolsep}{3pt}
\renewcommand{\arraystretch}{1.1}
\caption{Descriptive coverage of frequency-ranked static label sets.}
\label{tab:coverage}
\begin{tabular}{@{}rrrr@{}}
\toprule
\shortstack{Retained\\IDs} & \shortstack{Label-event\\coverage (\%)} & \shortstack{Complete\\OPs} & \shortstack{Complete\\OPs (\%)} \\
\midrule
12 & 74.6 & 0 & 0.00 \\
14 & 86.9 & 653 & 13.06 \\
15 & 91.3& 1,779 & 35.58 \\
17 & 97.8 & 3,374 & 67.48 \\
18 & 99.1 & 4,348 & 86.96 \\
19 & 99.9 & 4,997 & 99.94 \\
20 & 100.0 & 5,000 & 100.00 \\
\bottomrule
\end{tabular}
\par\smallskip
\end{table}
\subsection{{Two-Stage GAT Screening Performance}}

We first evaluate the trained two-stage GAT screener on the held-out test operating samples for the three severity budgets $\beta\in{0,0.1,0.3}$, each corresponding to a different MSS labelling regime. Table~\ref{tab:gat_summary} reports the results at the deliberately conservative operating threshold $\tau=0.005$. Here, ATB and APB denote the average numbers of true and predicted binding elements per test sample, respectively; TTB and TPB are their corresponding totals over the test set; TSC denotes the complete candidate space of 41 contingencies in StageA and 296 scalar constraints (per contingency in average) in StageB; and CovMu denotes the mean coverage of the true binding set.

The results demonstrate that the trained GAT can recover the security-relevant elements of the MSS from previously unseen operating samples while screening out a substantial portion of the candidate space. For Stage~A, the average number of predicted contingencies is 21.74, 19.07, and 16.02 for $\beta=0$, $0.1$, and $0.3$, respectively, compared with the complete set of 41 candidate contingencies. Thus, only approximately 48.3\%, 42.4\%, and 35.6\% of the contingency space needs to be retained, while a CovMu of 1.000 is achieved for all three severity budgets. The predicted sets are intentionally larger than the corresponding true sets (ATB), reflecting the conservative threshold selection: the model favors retaining additional contingencies rather than excluding security-relevant ones.

The screening capability is even more pronounced in StageB. From a candidate space of 296 scalar constraints, the model retains on average only 2.25, 2.48, and 1.46 constraints for $\beta=0$, $0.1$, and $0.3$, respectively. These values correspond to less than 1\% of the complete StageB candidate space, while CovMu remains 1.000 in all cases. Although the predicted Stage~B sets are larger than the true binding sets, particularly for $\beta=0$ and $\beta=0.1$, these additional predictions act conservatively by introducing extra constraints rather than omitting elements required by the reference MSS.

These results establish the practical applicability of the proposed two-stage screening architecture. Given an unseen pre-contingency operating state, StageA first restricts the security analysis to a substantially smaller subset of potentially relevant contingencies, after which StageB identifies a highly sparse subset of associated security constraints. Most importantly, the observed full coverage at the conservative operating threshold indicates that this reduction is achieved without excluding true MSS elements on the considered test set. The GAT therefore provides a data-driven approximation of the offline MSS construction procedure that can be directly used to define a compact reduced PSCOPF, avoiding the need to repeat the complete iterative security-screening procedure for every  $s$.
\begin{table}[t]
\centering
\caption{{Two-Stage GAT Screening Performance at the Conservative Operating Threshold $\tau=0.005$ for Severity Budgets $\beta\in\{0,0.1,0.3\}$}}
\label{tab:gat_summary}
\setlength{\tabcolsep}{4pt}

\begin{tabular}{llrrrrrr}
\toprule
Stage & $\beta$ & ATB & APB & TTB & TPB & TSC & CovMu \\
\midrule
Stage A & 0.0 & 18.12 & 21.74 & 4529 & 5435 & 45 & 1.000 \\
Stage B & 0.0 & 0.77 & 2.25 & 13895 & 40707 & 296 & 1.000 \\
\midrule
Stage A & 0.1 & 16.16 & 19.07 & 4040 & 4768 & 45 & 1.000 \\
Stage B & 0.1 & 0.78 & 2.48 & 12605 & 40125 & 296 & 1.000 \\
\midrule
Stage A & 0.3 & 13.41 & 16.02 & 3353 & 4005 & 45 & 1.000 \\
Stage B & 0.3 & 0.71 & 1.46 & 9576 & 19595 & 296 & 1.000 \\
\bottomrule
\end{tabular}
\end{table}

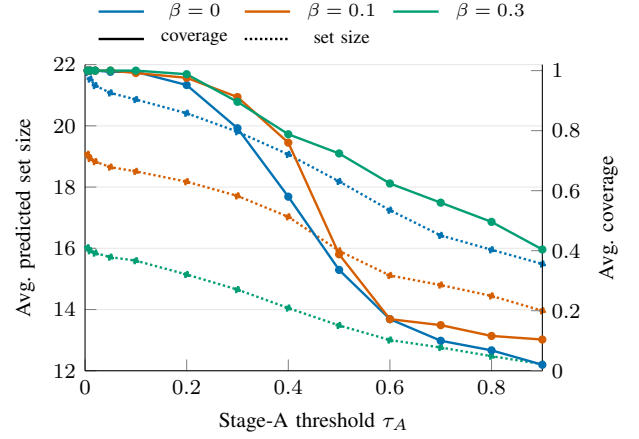
\begin{figure}[t]
\centering
\begin{tikzpicture}
\begin{axis}[
  width=6.05cm, height=4.05cm, scale only axis,
  xlabel={Stage-A threshold $\tau_A$}, xmin=0, xmax=0.9,
  axis y line*=left, axis x line*=bottom,
  ylabel={Avg.\ predicted set size}, ymin=12, ymax=22,
  ymajorgrids, grid style={gray!18},
  tick label style={font=\footnotesize},
  label style={font=\footnotesize},
  legend style={font=\scriptsize, at={(0.5,1.02)}, anchor=south,
    legend columns=3, draw=none, column sep=4pt},
]
\addlegendimage{siZero, line width=0.9pt}\addlegendentry{$\beta=0$}
\addlegendimage{siOne, line width=0.9pt}\addlegendentry{$\beta=0.1$}
\addlegendimage{siThree, line width=0.9pt}\addlegendentry{$\beta=0.3$}
\addlegendimage{black, solid, line width=0.9pt}\addlegendentry{coverage}
\addlegendimage{black, densely dotted, line width=0.9pt}\addlegendentry{set size}
\addplot[siZero, densely dotted, line width=0.9pt, mark=square*, mark size=0.9pt] coordinates {(0.005,21.74)(0.01,21.528)(0.02,21.316)(0.05,21.076)(0.1,20.86)(0.2,20.412)(0.3,19.808)(0.4,19.068)(0.5,18.184)(0.6,17.244)(0.7,16.42)(0.8,15.952)(0.9,15.488)};
\addplot[siOne, densely dotted, line width=0.9pt, mark=square*, mark size=0.9pt] coordinates {(0.005,19.072)(0.01,18.932)(0.02,18.832)(0.05,18.652)(0.1,18.512)(0.2,18.176)(0.3,17.712)(0.4,17.028)(0.5,15.916)(0.6,15.112)(0.7,14.8)(0.8,14.444)(0.9,13.96)};
\addplot[siThree, densely dotted, line width=0.9pt, mark=square*, mark size=0.9pt] coordinates {(0.005,16.02)(0.01,15.92)(0.02,15.844)(0.05,15.712)(0.1,15.596)(0.2,15.144)(0.3,14.652)(0.4,14.048)(0.5,13.48)(0.6,13.004)(0.7,12.76)(0.8,12.476)(0.9,12.216)};
\end{axis}
\begin{axis}[
  width=6.05cm, height=4.05cm, scale only axis,
  xmin=0, xmax=0.9, axis x line=none,
  axis y line*=right,
  ylabel={Avg.\ coverage}, ymin=0, ymax=1.02,
  tick label style={font=\footnotesize},
  label style={font=\footnotesize},
]
\addplot[siZero, solid, line width=0.9pt, mark=*, mark size=1.1pt] coordinates {(0.005,1)(0.01,1)(0.02,1)(0.05,0.996)(0.1,0.996)(0.2,0.952)(0.3,0.808)(0.4,0.58)(0.5,0.336)(0.6,0.172)(0.7,0.1)(0.8,0.068)(0.9,0.02)};
\addplot[siOne, solid, line width=0.9pt, mark=*, mark size=1.1pt] coordinates {(0.005,1)(0.01,1)(0.02,1)(0.05,1)(0.1,0.992)(0.2,0.976)(0.3,0.912)(0.4,0.76)(0.5,0.388)(0.6,0.172)(0.7,0.152)(0.8,0.116)(0.9,0.104)};
\addplot[siThree, solid, line width=0.9pt, mark=*, mark size=1.1pt] coordinates {(0.005,1)(0.01,1)(0.02,1)(0.05,1)(0.1,1)(0.2,0.988)(0.3,0.896)(0.4,0.788)(0.5,0.724)(0.6,0.624)(0.7,0.56)(0.8,0.496)(0.9,0.404)};
\end{axis}
\end{tikzpicture}
\caption{{Stage-A (contingency screening) trade-off between coverage (solid, right axis) and average predicted contingency-set size (dotted, left axis) as the decision threshold $\tau_A$ is swept, for the three severity budgets $\beta\in\{0,0.1,0.3\}$.}}
\label{fig:stageA_tau}
\end{figure}

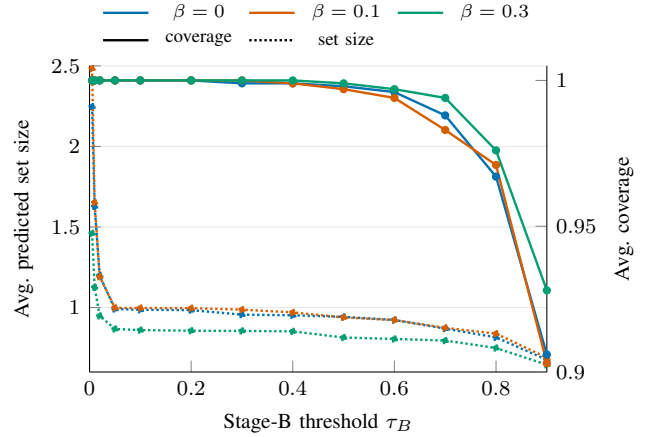
\begin{figure}[t]
\centering
\begin{tikzpicture}
\begin{axis}[
  width=6.05cm, height=4.05cm, scale only axis,
  xlabel={Stage-B threshold $\tau_B$}, xmin=0, xmax=0.9,
  axis y line*=left, axis x line*=bottom,
  ylabel={Avg.\ predicted set size}, ymin=0.6, ymax=2.5,
  ymajorgrids, grid style={gray!18},
  tick label style={font=\footnotesize},
  label style={font=\footnotesize},
  legend style={font=\scriptsize, at={(0.5,1.02)}, anchor=south,
    legend columns=3, draw=none, column sep=4pt},
]
\addlegendimage{siZero, line width=0.9pt}\addlegendentry{$\beta=0$}
\addlegendimage{siOne, line width=0.9pt}\addlegendentry{$\beta=0.1$}
\addlegendimage{siThree, line width=0.9pt}\addlegendentry{$\beta=0.3$}
\addlegendimage{black, solid, line width=0.9pt}\addlegendentry{coverage}
\addlegendimage{black, densely dotted, line width=0.9pt}\addlegendentry{set size}
\addplot[siZero, densely dotted, line width=0.9pt, mark=square*, mark size=0.9pt] coordinates {(0.005,2.247)(0.01,1.628)(0.02,1.198)(0.05,0.991)(0.1,0.986)(0.2,0.984)(0.3,0.956)(0.4,0.952)(0.5,0.942)(0.6,0.923)(0.7,0.869)(0.8,0.815)(0.9,0.676)};
\addplot[siOne, densely dotted, line width=0.9pt, mark=square*, mark size=0.9pt] coordinates {(0.005,2.483)(0.01,1.652)(0.02,1.191)(0.05,0.998)(0.1,0.997)(0.2,0.996)(0.3,0.987)(0.4,0.97)(0.5,0.939)(0.6,0.922)(0.7,0.874)(0.8,0.838)(0.9,0.689)};
\addplot[siThree, densely dotted, line width=0.9pt, mark=square*, mark size=0.9pt] coordinates {(0.005,1.461)(0.01,1.126)(0.02,0.948)(0.05,0.867)(0.1,0.86)(0.2,0.856)(0.3,0.855)(0.4,0.852)(0.5,0.814)(0.6,0.805)(0.7,0.795)(0.8,0.749)(0.9,0.645)};
\end{axis}
\begin{axis}[
  width=6.05cm, height=4.05cm, scale only axis,
  xmin=0, xmax=0.9, axis x line=none,
  axis y line*=right,
  ylabel={Avg.\ coverage}, ymin=0.9, ymax=1.005,
  tick label style={font=\footnotesize},
  label style={font=\footnotesize},
]
\addplot[siZero, solid, line width=0.9pt, mark=*, mark size=1.1pt] coordinates {(0.005,1)(0.01,1)(0.02,1)(0.05,1)(0.1,1)(0.2,1)(0.3,0.999)(0.4,0.999)(0.5,0.998)(0.6,0.996)(0.7,0.988)(0.8,0.967)(0.9,0.906)};
\addplot[siOne, solid, line width=0.9pt, mark=*, mark size=1.1pt] coordinates {(0.005,1)(0.01,1)(0.02,1)(0.05,1)(0.1,1)(0.2,1)(0.3,1)(0.4,0.999)(0.5,0.997)(0.6,0.994)(0.7,0.983)(0.8,0.971)(0.9,0.903)};
\addplot[siThree, solid, line width=0.9pt, mark=*, mark size=1.1pt] coordinates {(0.005,1)(0.01,1)(0.02,1)(0.05,1)(0.1,1)(0.2,1)(0.3,1)(0.4,1)(0.5,0.999)(0.6,0.997)(0.7,0.994)(0.8,0.976)(0.9,0.928)};
\end{axis}
\end{tikzpicture}
\caption{{Stage-B (constraint screening) trade-off between coverage (solid, right axis) and average predicted constraint-set size (dotted, left axis) versus the decision threshold $\tau_B$, for the three severity budgets $\beta\in\{0,0.1,0.3\}$.}}
\label{fig:stageB_tau}
\end{figure}

{The threshold sweeps in Figs.~\ref{fig:stageA_tau} and \ref{fig:stageB_tau} expose the coverage/size trade-off that these operating points sit on. For both stages the coverage curves (solid, right axis) remain essentially at unity over a wide low-threshold band while the predicted-set-size curves (dotted, left axis) decrease steadily, before coverage drops sharply once the threshold passes a knee. This flat region is exactly what an operator wants: the thresholds can be pushed to trim the retained set while the binding elements are still (almost) fully captured. Larger severity budgets shift both families of curves downward, because a looser budget labels fewer contingencies and constraints as binding and therefore yields an intrinsically smaller MSS to reproduce.}

\begin{figure}[t]
\centering
\begin{tikzpicture}
\begin{axis}[
  width=\columnwidth, height=0.62\columnwidth,
  ymode=log,
  xlabel={Epoch}, ylabel={Stage-B loss (log)},
  xmin=1, xmax=50,
  ymajorgrids, grid style={gray!18},
  tick label style={font=\footnotesize},
  label style={font=\footnotesize},
  legend style={font=\scriptsize, at={(0.98,0.98)}, anchor=north east,
    draw=none, fill=white, fill opacity=0.7, text opacity=1, legend columns=2, column sep=3pt},
]
\addplot[siZero, solid, line width=0.9pt] coordinates {(1,0.9336)(2,0.1242)(3,0.0698)(4,0.0553)(5,0.0505)(6,0.0481)(7,0.0467)(8,0.046)(9,0.045)(10,0.0446)(11,0.0437)(12,0.0435)(13,0.043)(14,0.0425)(15,0.0422)(16,0.042)(17,0.0418)(18,0.0418)(19,0.0415)(20,0.0414)(21,0.0413)(22,0.0412)(23,0.0411)(24,0.0408)(25,0.0409)(26,0.0407)(27,0.0407)(28,0.0405)(29,0.0403)(30,0.0404)(31,0.0403)(32,0.0402)(33,0.04)(34,0.0399)(35,0.04)(36,0.0398)(37,0.0397)(38,0.0397)(39,0.0396)(40,0.0397)(41,0.0395)(42,0.0395)(43,0.0395)(44,0.0395)(45,0.0393)(46,0.0394)(47,0.0394)(48,0.0392)(49,0.0393)(50,0.0391)};
\addlegendentry{$\beta=0$ train}
\addplot[siOne, solid, line width=0.9pt] coordinates {(1,0.9377)(2,0.1253)(3,0.0648)(4,0.0541)(5,0.0501)(6,0.0481)(7,0.0469)(8,0.0461)(9,0.0455)(10,0.045)(11,0.0445)(12,0.0442)(13,0.0435)(14,0.0436)(15,0.0432)(16,0.0429)(17,0.0427)(18,0.0427)(19,0.0424)(20,0.0423)(21,0.0421)(22,0.0421)(23,0.0421)(24,0.0421)(25,0.0419)(26,0.0419)(27,0.0417)(28,0.0419)(29,0.0417)(30,0.0416)(31,0.0415)(32,0.0414)(33,0.0413)(34,0.0413)(35,0.0414)(36,0.0411)(37,0.0412)(38,0.0412)(39,0.0411)(40,0.041)(41,0.0409)(42,0.041)(43,0.041)(44,0.0409)(45,0.0409)(46,0.0411)(47,0.0406)(48,0.0406)(49,0.0406)(50,0.0408)};
\addlegendentry{$\beta=0.1$ train}
\addplot[siThree, solid, line width=0.9pt] coordinates {(1,0.8914)(2,0.0877)(3,0.0446)(4,0.0385)(5,0.0366)(6,0.0355)(7,0.0345)(8,0.0338)(9,0.0333)(10,0.0327)(11,0.0323)(12,0.0319)(13,0.0317)(14,0.0315)(15,0.0311)(16,0.0311)(17,0.031)(18,0.0308)(19,0.0306)(20,0.0305)(21,0.0305)(22,0.0305)(23,0.0305)(24,0.0301)(25,0.0303)(26,0.0305)(27,0.0302)(28,0.03)(29,0.0299)(30,0.03)(31,0.0299)(32,0.0298)(33,0.03)(34,0.0299)(35,0.0297)(36,0.0297)(37,0.0295)(38,0.0294)(39,0.0298)(40,0.0294)(41,0.0298)(42,0.0294)(43,0.0294)(44,0.0295)(45,0.0296)(46,0.0296)(47,0.0297)(48,0.0294)(49,0.0291)(50,0.0294)};
\addlegendentry{$\beta=0.3$ train}
\addplot[siZero, densely dotted, line width=0.9pt] coordinates {(1,0.1656)(2,0.064)(3,0.0461)(4,0.0406)(5,0.0394)(6,0.0386)(7,0.0382)(8,0.0386)(9,0.038)(10,0.0383)(11,0.0377)(12,0.0378)(13,0.0367)(14,0.0365)(15,0.0371)(16,0.0365)(17,0.0364)(18,0.0368)(19,0.0359)(20,0.0364)(21,0.0363)(22,0.0362)(23,0.0361)(24,0.0361)(25,0.0356)(26,0.0356)(27,0.0362)(28,0.0357)(29,0.0352)(30,0.0351)(31,0.0358)(32,0.0348)(33,0.0352)(34,0.0356)(35,0.035)(36,0.0357)(37,0.0362)(38,0.0353)(39,0.0355)(40,0.0357)(41,0.0352)(42,0.0353)(43,0.0349)(44,0.0358)(45,0.035)(46,0.0352)(47,0.0348)(48,0.0348)(49,0.035)(50,0.0352)};
\addlegendentry{$\beta=0$ val.}
\addplot[siOne, densely dotted, line width=0.9pt] coordinates {(1,0.1765)(2,0.0588)(3,0.0453)(4,0.0421)(5,0.0413)(6,0.0413)(7,0.0412)(8,0.0409)(9,0.0401)(10,0.0401)(11,0.0403)(12,0.0395)(13,0.04)(14,0.0397)(15,0.0392)(16,0.039)(17,0.0395)(18,0.0386)(19,0.0386)(20,0.0386)(21,0.0385)(22,0.039)(23,0.0385)(24,0.0387)(25,0.0386)(26,0.0395)(27,0.0384)(28,0.0387)(29,0.0383)(30,0.0394)(31,0.0383)(32,0.0387)(33,0.0386)(34,0.0393)(35,0.0385)(36,0.0382)(37,0.0385)(38,0.038)(39,0.0382)(40,0.0384)(41,0.038)(42,0.039)(43,0.0381)(44,0.0378)(45,0.0407)(46,0.0383)(47,0.039)(48,0.0384)(49,0.0387)(50,0.0393)};
\addlegendentry{$\beta=0.1$ val.}
\addplot[siThree, densely dotted, line width=0.9pt] coordinates {(1,0.1364)(2,0.0373)(3,0.0314)(4,0.0299)(5,0.0301)(6,0.0297)(7,0.0288)(8,0.0288)(9,0.0279)(10,0.0293)(11,0.0275)(12,0.0277)(13,0.0273)(14,0.029)(15,0.0271)(16,0.0277)(17,0.0282)(18,0.0275)(19,0.0271)(20,0.0272)(21,0.0273)(22,0.027)(23,0.027)(24,0.0279)(25,0.0266)(26,0.0274)(27,0.0267)(28,0.0268)(29,0.0273)(30,0.0269)(31,0.027)(32,0.0279)(33,0.0274)(34,0.0281)(35,0.0277)(36,0.0271)(37,0.0266)(38,0.0286)(39,0.0269)(40,0.0281)(41,0.027)(42,0.0275)(43,0.0294)(44,0.0271)(45,0.0272)(46,0.0285)(47,0.0276)(48,0.027)(49,0.0269)(50,0.0277)};
\addlegendentry{$\beta=0.3$ val.}
\end{axis}
\end{tikzpicture}
\caption{{Stage-B training loss (solid) and validation loss (dotted) versus epoch for the three severity budgets $\beta\in\{0,0.1,0.3\}$ (logarithmic vertical axis).}}
\label{fig:lossB}
\end{figure}
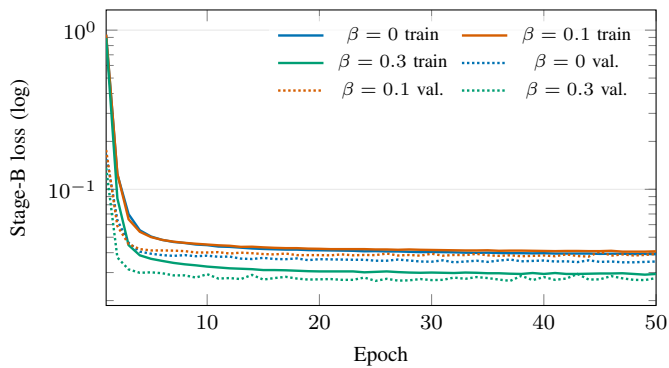

{The learning curves in Figs.~\ref{fig:lossA} and \ref{fig:lossB} confirm that this behaviour reflects a well-trained model rather than an artefact of thresholding. For every severity budget the Stage-A and Stage-B losses fall quickly and then flatten, and the validation curves (dotted) track the training curves (solid) closely with no divergence, indicating that the GAT generalises to unseen operating points without overfitting. The Stage-B loss in particular collapses within the first few epochs from its large initial value, reflecting how sparse and structured the binding-constraint labels are once the correct contingencies are identified.}

\begin{figure}[t]
\centering
\begin{tikzpicture}
\begin{axis}[
  width=\columnwidth, height=0.62\columnwidth,
  ymode=log,
  xlabel={Epoch}, ylabel={Stage-A loss (log)},
  xmin=1, xmax=50,
  ymajorgrids, grid style={gray!18},
  tick label style={font=\footnotesize},
  label style={font=\footnotesize},
  legend style={font=\scriptsize, at={(0.98,0.98)}, anchor=north east,
    draw=none, fill=white, fill opacity=0.7, text opacity=1, legend columns=2, column sep=3pt},
]
\addplot[siZero, solid, line width=0.9pt] coordinates {(1,0.171)(2,0.1239)(3,0.1229)(4,0.1214)(5,0.1219)(6,0.1201)(7,0.1182)(8,0.1133)(9,0.1075)(10,0.1035)(11,0.1008)(12,0.0987)(13,0.0987)(14,0.0971)(15,0.0955)(16,0.095)(17,0.0938)(18,0.0951)(19,0.0933)(20,0.0919)(21,0.0929)(22,0.0914)(23,0.0925)(24,0.0939)(25,0.0919)(26,0.0903)(27,0.09)(28,0.0897)(29,0.0902)(30,0.0902)(31,0.0909)(32,0.09)(33,0.0882)(34,0.0891)(35,0.0901)(36,0.0889)(37,0.0894)(38,0.0884)(39,0.0869)(40,0.089)(41,0.0892)(42,0.0898)(43,0.0896)(44,0.0891)(45,0.0902)(46,0.0868)(47,0.0907)(48,0.0884)(49,0.0871)(50,0.0882)};
\addlegendentry{$\beta=0$ train}
\addplot[siOne, solid, line width=0.9pt] coordinates {(1,0.1475)(2,0.0951)(3,0.0946)(4,0.0939)(5,0.0919)(6,0.0902)(7,0.0874)(8,0.0859)(9,0.0841)(10,0.0825)(11,0.0817)(12,0.0811)(13,0.0803)(14,0.0799)(15,0.0797)(16,0.0786)(17,0.0794)(18,0.0791)(19,0.0791)(20,0.0789)(21,0.0786)(22,0.0784)(23,0.0777)(24,0.0792)(25,0.0782)(26,0.0776)(27,0.0775)(28,0.0777)(29,0.0784)(30,0.0789)(31,0.0776)(32,0.0771)(33,0.0771)(34,0.0779)(35,0.0779)(36,0.0767)(37,0.077)(38,0.0771)(39,0.0781)(40,0.0777)(41,0.0769)(42,0.077)(43,0.077)(44,0.077)(45,0.0762)(46,0.0768)(47,0.0764)(48,0.0762)(49,0.0768)(50,0.0766)};
\addlegendentry{$\beta=0.1$ train}
\addplot[siThree, solid, line width=0.9pt] coordinates {(1,0.1488)(2,0.0884)(3,0.0869)(4,0.0858)(5,0.0858)(6,0.0835)(7,0.0803)(8,0.0783)(9,0.0765)(10,0.0749)(11,0.0724)(12,0.0727)(13,0.0725)(14,0.0715)(15,0.0711)(16,0.0704)(17,0.07)(18,0.0696)(19,0.069)(20,0.0694)(21,0.0701)(22,0.0688)(23,0.0693)(24,0.0684)(25,0.0681)(26,0.07)(27,0.0693)(28,0.0685)(29,0.0686)(30,0.0688)(31,0.0685)(32,0.0685)(33,0.069)(34,0.0674)(35,0.0686)(36,0.0679)(37,0.0672)(38,0.0681)(39,0.068)(40,0.0671)(41,0.0674)(42,0.0675)(43,0.0679)(44,0.068)(45,0.0675)(46,0.067)(47,0.0681)(48,0.0671)(49,0.0677)(50,0.0669)};
\addlegendentry{$\beta=0.3$ train}
\addplot[siZero, densely dotted, line width=0.9pt] coordinates {(1,0.1163)(2,0.1118)(3,0.1126)(4,0.1109)(5,0.1142)(6,0.1087)(7,0.1072)(8,0.1006)(9,0.0946)(10,0.091)(11,0.0947)(12,0.0993)(13,0.0914)(14,0.0891)(15,0.093)(16,0.0887)(17,0.0872)(18,0.0986)(19,0.0885)(20,0.0873)(21,0.0887)(22,0.0863)(23,0.0869)(24,0.0864)(25,0.0871)(26,0.0865)(27,0.0861)(28,0.0869)(29,0.0851)(30,0.0856)(31,0.0849)(32,0.0847)(33,0.0851)(34,0.0848)(35,0.0849)(36,0.0857)(37,0.0905)(38,0.0891)(39,0.0855)(40,0.0862)(41,0.0854)(42,0.0904)(43,0.0846)(44,0.0856)(45,0.084)(46,0.0835)(47,0.0853)(48,0.0846)(49,0.0859)(50,0.0843)};
\addlegendentry{$\beta=0$ val.}
\addplot[siOne, densely dotted, line width=0.9pt] coordinates {(1,0.0895)(2,0.09)(3,0.0865)(4,0.0866)(5,0.0853)(6,0.0826)(7,0.0838)(8,0.0786)(9,0.0754)(10,0.0779)(11,0.0751)(12,0.0739)(13,0.0746)(14,0.0722)(15,0.073)(16,0.0715)(17,0.0746)(18,0.0739)(19,0.0733)(20,0.0716)(21,0.0715)(22,0.0754)(23,0.0714)(24,0.0721)(25,0.0738)(26,0.0723)(27,0.072)(28,0.0713)(29,0.0715)(30,0.0713)(31,0.0709)(32,0.0709)(33,0.0725)(34,0.0718)(35,0.0717)(36,0.0718)(37,0.0711)(38,0.0707)(39,0.0704)(40,0.0709)(41,0.0708)(42,0.0717)(43,0.0722)(44,0.0721)(45,0.0712)(46,0.0717)(47,0.0708)(48,0.071)(49,0.0716)(50,0.071)};
\addlegendentry{$\beta=0.1$ val.}
\addplot[siThree, densely dotted, line width=0.9pt] coordinates {(1,0.0928)(2,0.0913)(3,0.0898)(4,0.095)(5,0.0889)(6,0.0883)(7,0.082)(8,0.0784)(9,0.0752)(10,0.075)(11,0.0713)(12,0.0719)(13,0.0714)(14,0.0729)(15,0.069)(16,0.0705)(17,0.0722)(18,0.0685)(19,0.0717)(20,0.0719)(21,0.0716)(22,0.0697)(23,0.067)(24,0.0714)(25,0.067)(26,0.0686)(27,0.0661)(28,0.067)(29,0.0667)(30,0.0679)(31,0.0674)(32,0.0659)(33,0.0662)(34,0.0711)(35,0.0663)(36,0.0655)(37,0.0679)(38,0.0703)(39,0.0671)(40,0.0651)(41,0.066)(42,0.0669)(43,0.0678)(44,0.0693)(45,0.072)(46,0.066)(47,0.0658)(48,0.065)(49,0.0651)(50,0.0668)};
\addlegendentry{$\beta=0.3$ val.}
\end{axis}
\end{tikzpicture}
\caption{{Stage-A training loss (solid) and validation loss (dotted) versus epoch for the three severity budgets $\beta\in\{0,0.1,0.3\}$ (logarithmic vertical axis).}}
\label{fig:lossA}
\end{figure}
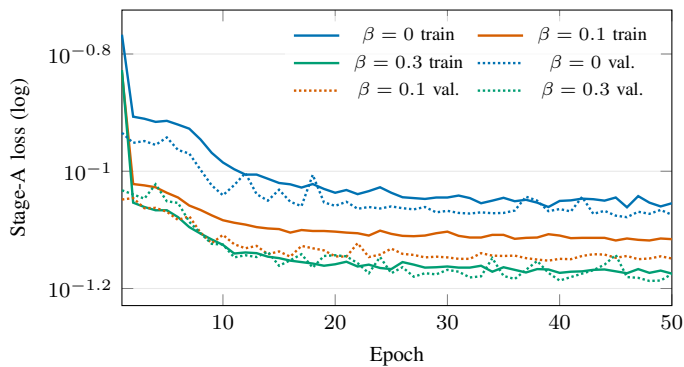

{Taken together, these results show that the two-stage screener shrinks the severity-aware AC-PSCOPF very aggressively while keeping the security-relevant elements. At the conservative $\beta=0$ setting of Table~\ref{tab:gat_summary}, Stage~A retains on average only $21.7$ of the $45$ candidate contingencies per sample---already discarding about half of them---while still covering every truly binding contingency (CovMu $=1.0$). Stage~B then acts on each retained contingency and keeps roughly two of its $296$ monitored limits, so that the predicted binding-constraint set spans about $40.7$k scalar slots against the $3.33$M slots of the full monitored space over the test set, a reduction of close to $99\%$ at full coverage. Because binding contingencies are identified \emph{before} the optimisation is assembled, the reduced PSCOPF that an operator must actually solve is one to two orders of magnitude smaller than the full problem, and the near-unit coverage guarantees that the limits that would shape the AC-PSCOPF solution are the very ones retained. As the severity budget is relaxed ($\beta=0.1,0.3$), both the true and predicted sets shrink further, confirming that the framework adapts the size of the security representation to the tolerated risk level without sacrificing coverage.}

\section{Conclusion}
This paper proposed a severity-aware ML-assisted framework for identifying minimum security sets in preventive SCOPF. The proposed severity criterion extends conventional binary security screening by accounting for the number, accumulated magnitude, maximum magnitude, and average magnitude of post-contingency violations. An offline MSS construction procedure was developed by combining AC security assessment, non-dominated contingency screening, and iterative reduced PSCOPF. The resulting MSS labels can then be used to train a supervised multi-label model to predict the relevant contingency--constraint pairs directly from the pre-contingency operating state.

The framework was evaluated on the IEEE 60-bus and IEEE 118-bus test systems, demonstrating substantial reductions in computational burden while maintaining a preventive operating point close to that of the full PSCOPF. For the 60-bus system, the highest severity level reduced the number of variables and constraints by $66.7\%$ and $71.5\%$, respectively, and reduced solution time by $82.8\%$. For the larger IEEE 118-bus system, up to $57.8\%$ of the contingencies were removed, leading to reductions of $39.2\%$ in variables and $48.4\%$ in constraints and an $89.5\%$ reduction in solution time. These results show that contingency and constraint filtering substantially reduce computation, particularly for larger PSCOPF problems.

The severity threshold provides a direct mechanism for controlling the trade-off between computational efficiency and residual post-contingency violations. The exhaustive post-optimization security assessment retains visibility over the complete contingency set and allows any violations exceeding the prescribed severity threshold to be added back to the reduced PSCOPF if required. Future work will focus on larger-scale systems, alternative ML architectures, and online validation of the learned MSS prediction for operational deployment.

\bibliographystyle{IEEEtran}
\bibliography{refs}

\end{document}